\documentclass[english,final]{siamltex1213}

\usepackage{graphicx}
\usepackage{verbatim}
\usepackage{amssymb}
\usepackage{graphicx}
\usepackage{color}
\usepackage{wrapfig}
\usepackage{caption}
\usepackage{subcaption}

\makeatletter

\newcommand\eqref[1]{(\ref{#1})}

\usepackage{babel}

\makeatother

\title{Sensitivity analysis of neuronal dynamics based on additive STDP rule}
\author{Subhajit Sengupta\footnotemark[3] 
\and Karthik S. Gurumoorthy\footnotemark[3] 
\and Arunava Banerjee \footnotemark[3]
}
\begin{document}
\maketitle
\renewcommand{\thefootnote}{\fnsymbol{footnote}}
\footnotetext[3]{Department of Computer and Information Science and Engineering,
University of Florida, Gainesville, Florida, USA. \\ Emails: {\tt subhajit06@gmail.com}, {\tt karthik.gurumoorthy@gmail.com}, {\tt arunava@cise.ufl.edu}}

\begin{abstract}
Spike Timing Dependent Plasticity (STDP) is a Hebbian like synaptic learning rule. The basis of STDP has strong experimental evidences and it depends on precise input and output spike timings.  In this paper we show that under biologically plausible spiking regime, slight variability in the spike timing leads to drastically different evolution of synaptic weights when its dynamics are governed by the additive STDP rule. 
\end{abstract}

\pagestyle{myheadings}
\thispagestyle{plain}

\section{Introduction}
Spike Timing Dependent Plasticity (STDP) \cite{song2000competitive,abbott2000synaptic,van2000stable, markram1997regulation,bi1998synaptic,dan2004spike, caporale2008spike} is a temporally asymmetric form of Hebbian learning which has been experimentally observed in a wide spectrum of species. It is widely believed that STDP plays a key role in learning \cite{abbott2004homeostasis,gerstner1996neuronal}, information storage \cite{lengyel2005matching}, coincidence detection \cite{karmarkar2002model} as well as the  neuronal circuits development in the brain. STDP rule also has attractive features such as synaptic competition and rate stabilization \cite{song2000competitive}. It is induced by the strong temporal correlation between the spike timings of presynaptic and postsynaptic neurons. The synaptic modification depends on the precise timing and the order of the input and output spikes. The synaptic connection between two neurons is more likely to strengthen if the presynaptic neuron fires shortly before the postsynaptic neuron and similarly weakens if the inverse situation happens \cite{pfister2006optimal}. As STDP depends on specific spike timing, it would be interesting to investigate a scenario where those spike timings are perturbed slightly. In other words, we seek to answer the question: \emph{``Is STDP  robust enough to make synaptic weights insensitive to slight variations in input (or output) spike timings?''} Our analysis reveals the contrary where small initial perturbations have drastic effects on synaptic weights when its dynamics are governed by additive STDP rule. We would like to accentuate that though the fragility of the additive STDP rule concerned with the stabilization issues has been previously reported in the literature \cite{kepecs2002neuronal,van2000stable,rubin2001equilibrium}, our analysis is different and directed from a dynamical system viewpoint.

In order to investigate this, we studied the dynamics of a single output neuron with multiple synapses whose weight changes are dictated by the STDP rule. At the initial STDP time-window we give a tiny perturbation on input or output or both spikes. Later we do not allow any other perturbations in this system and let the system evolve in time. Our formal study shows that under biologically plausible spiking regimes the evolution of synaptic weights would be very different in this perturbed case than the original unperturbed system due to this initial small perturbations. For our analysis, we framed an equivalent problem to this by fixing the spike timings in the initial STDP time-window and adding a small perturbation on the synaptic weight itself. We argue that this two situations are exactly same because due to those small perturbations in input or output spike timings, the STDP update for synaptic weight would be slightly different in both cases, which will serve as the initial synaptic weight perturbation. In the entire analysis, we work with the perturbation vector associated with synaptic weight and prove our results. Our analysis considers a spiking neuron as an abstract dynamical system as described in \cite{banerjee2006sensitive}. In \cite{banerjee2006sensitive} the dynamical system was studied in the context of recurrent spiking neuron network assuming a fixed synaptic weight. In our case we study the dynamics of a single neuron particularly focusing on how STDP rule on excitatory synapses affects initial perturbations associated with synaptic weights. We study the effects of the six events, namely ($1$) \emph{birth of an output spike}, ($2$) \emph{birth of an input spike}, ($3$) \emph{death of an output spike}, ($4$) \emph{death of an input spike}, ($5$) \emph{ceiling} and ($6$) \emph{floor}. The last two events are the manifestation of the additive STDP rule.

\subsection{A simple experiment}
We start with the following simple experiment. Consider a single output neuron innervated by $1000$ input synapses among which $20\%$ are inhibitory. Spiking dynamics in the system are governed by the first-order Spike Response Model \cite{gerstner2002spiking}. At each timestep, the output neuron’s membrane potential is computed as the sum of all Post-Synaptic Potentials (PSP) of its input spikes, summed with the After Hyper-Polarizing (AHP) effects of previous output spikes. Form of PSP is given by $P_i(x_{i}^{k_j}) = \frac{1}{d\sqrt{x_{i}^{k_j}}}e^{-\frac{\beta d^{2}}{x_{i}^{k_j}}} \, e^{\frac{x_{i}^{k_j}}{\gamma_{P}}}$
where $x_{i}^{k_j}$ denotes the elapsed time at the $i^{th}$ synapse since the occurrence of the event $k_j$ (the notion of an event is made clear below), $d(>0)$ denotes the distance (in dimensionless units) of the synapse from the soma, and $\beta(>0)$ and $\gamma_P(>0)$ control the rate of rise and fall of the PSP. The after hyper-polarizing potential (AHP) effects are determined completely by the time since the departure of each output spike. The AHP is given by $P_0(x_{0}^{k_j}) = R e^{\frac{-x_{0}^{k_j}}{\gamma_A}} $ where $R(< 0)$ denotes the instantaneous fall in potential after a spike and $\gamma_A(> 0)$ controls its rate of recovery. $x_{0}^{k_j}$ represents the elapsed time since the event $k_j$ took place. In all experiments, we set $\beta = 6$ (dimensionless), $\tau = 15msec$, $R = -1000mV$ and $ \gamma_A= 1.6msec$. For additive STDP, $A_{+}$ and $A_{-}$ determine the maximum synaptic update amounts, and $\tau_{+}$ and $\tau_{-}$ determine the steepness of the exponential response curves. We use $A_{+} = 0.01$, $A_{-} = 0.0105$ and $\tau_{+} = \tau_{-} = 20msec$. The synaptic weights between inputs and output are initialized randomly between $w_{max}$ and $w_{min}$ whose values are $40$ and $0$ respectively. 

\begin{figure}
        \begin{subfigure}[b]{0.31\textwidth}
                \centering
                \includegraphics[scale=0.32]{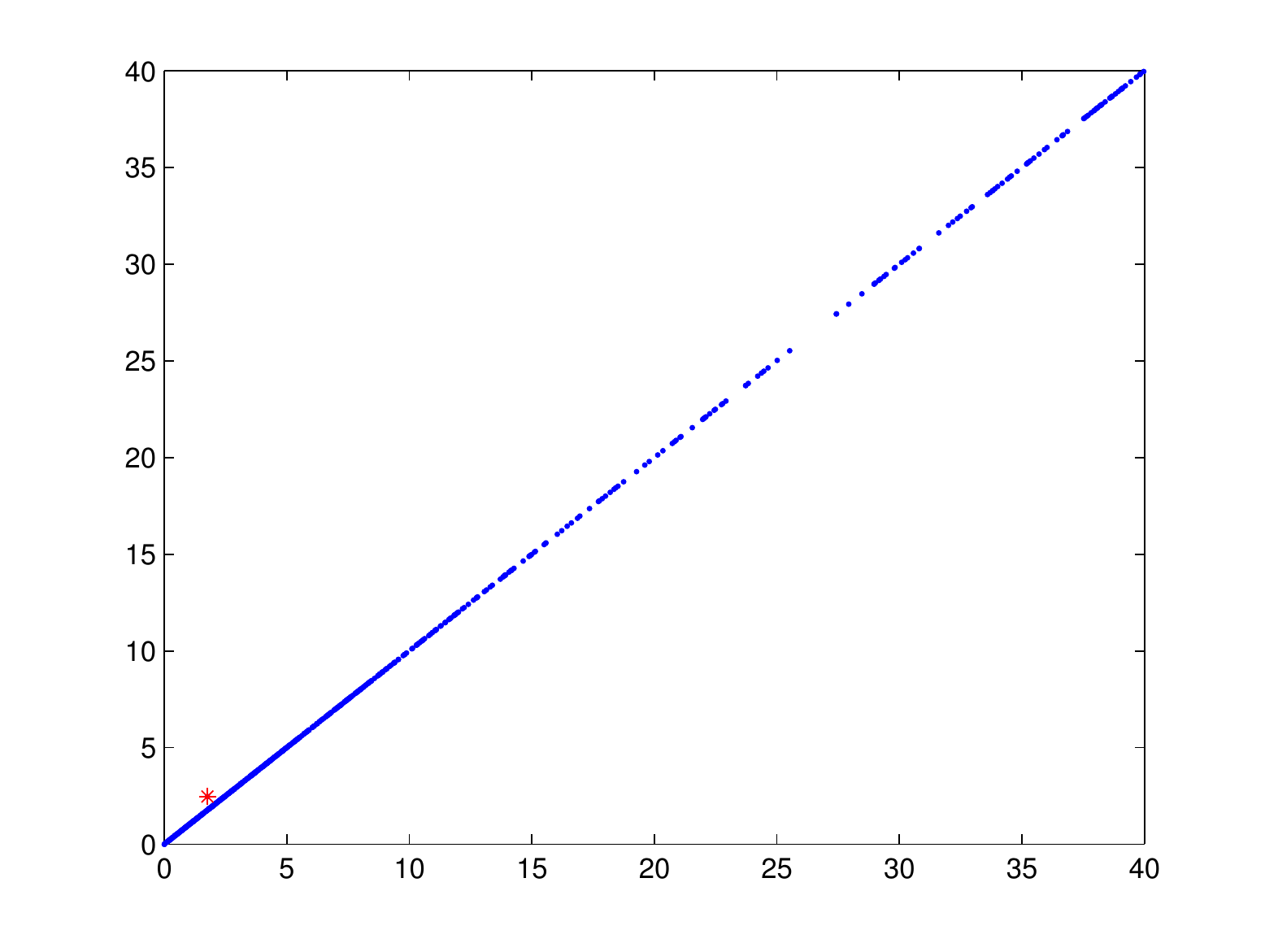}
                \vspace{-0.5pt}
                \caption{Both systems - Initial}
                \label{fig:init_perturbation}
        \end{subfigure}
        ~ 
        \begin{subfigure}[b]{0.31\textwidth}
                \centering
                \includegraphics[scale=0.32]{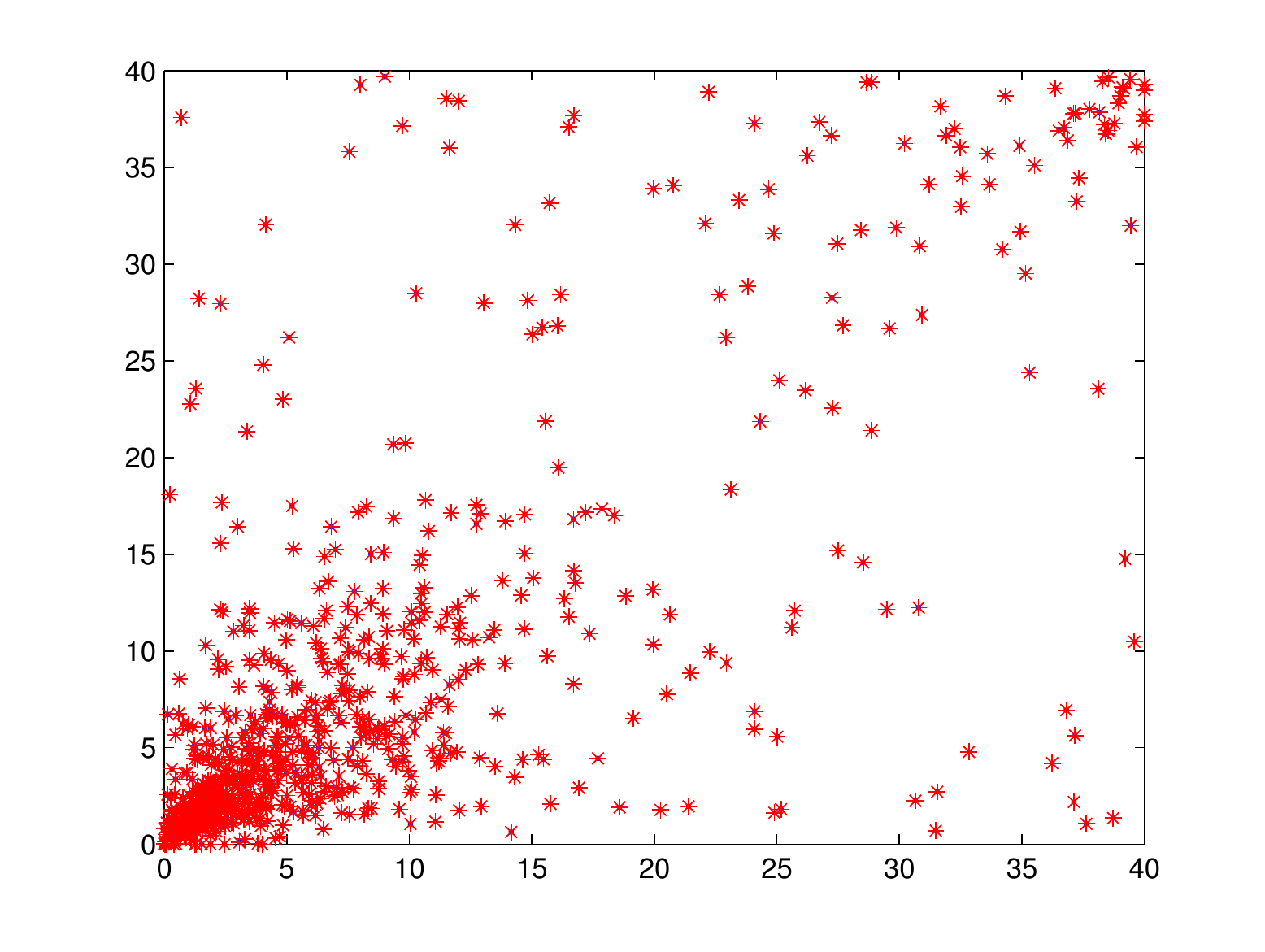}
                \vspace{-1pt}
                \caption{Both systems - Final}
                \label{fig:final_perturbation}
        \end{subfigure}
        ~ 
        \begin{subfigure}[b]{0.31\textwidth}
                \centering
                \includegraphics[scale=0.335]{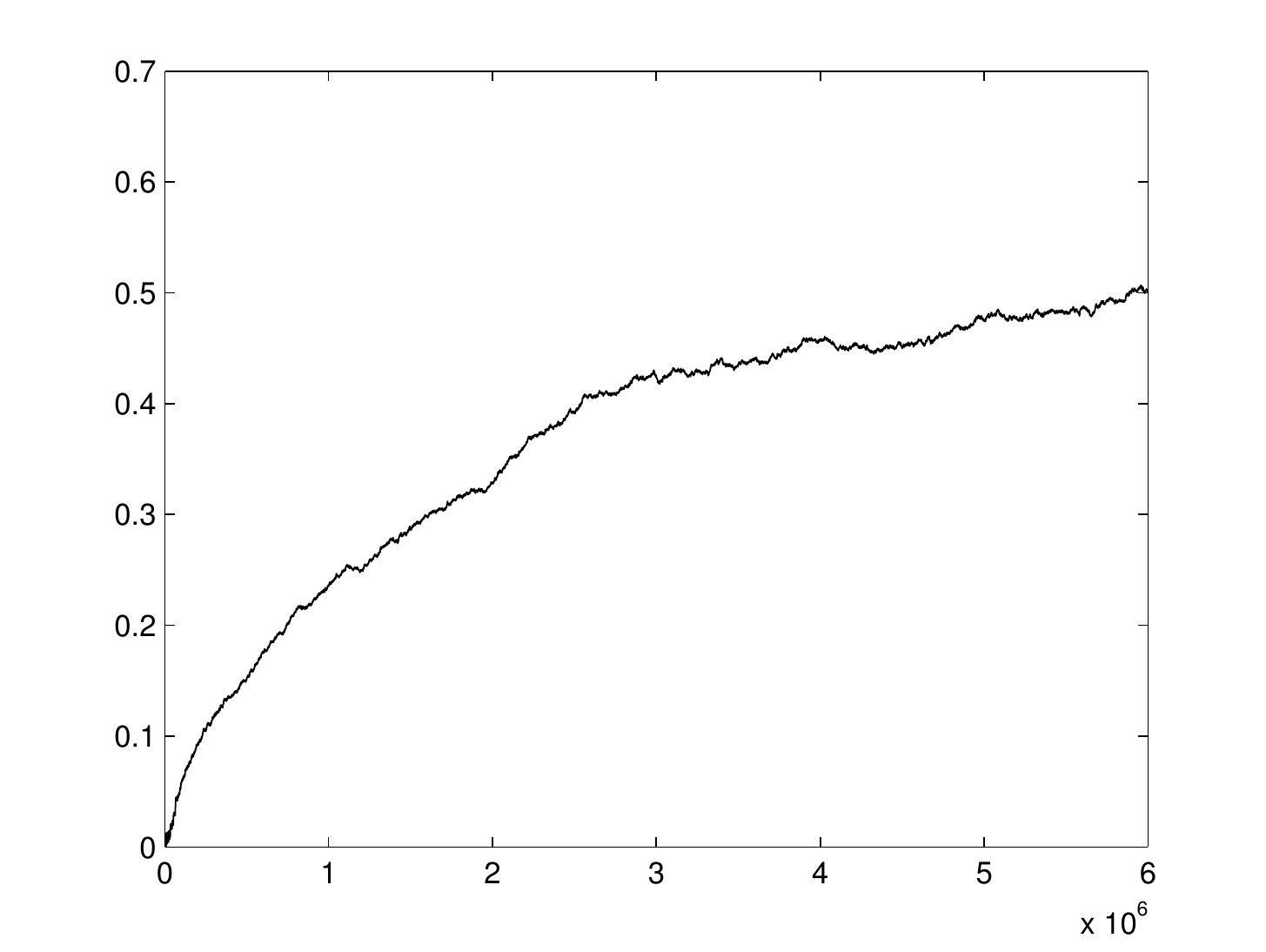}           
                \caption{Normalized sum of squared differences}
                \label{fig:sq_diff}
        \end{subfigure}
        \caption{}
        \label{fig:perturbation}
\end{figure}

 We simulate two runs of the same system with identical Poisson spike train inputs.  We run them simultaneously till the synaptic weights reach to a steady state distribution. Once the weights have stabilized, we slightly perturb the weight of a single synapse in one of the system. In the Figure~\ref{fig:init_perturbation}, we show the 2D plot of the synaptic weights for the perturbed and the unperturbed system in the \textit{x-axis} and \textit{y-axis} respectively and they are same except for a single synapse (the red dot shifted away from the diagonal corresponds to the perturbed synapse). We then let both the perturbed and the unperturbed systems to evolve with the synaptic weights governed by the additive STDP rule. After some long enough time, final synaptic weights for both the systems are noted and plotted in the Figure~\ref{fig:final_perturbation}. Equal synaptic weight in both the systems are shown in `blue' and synapses with different weights in both the systems are shown in `red'. We detect that it has been deviated from the $x=y$ line for almost all the places - which clearly indicates sensitivity to initial condition. The claim is further strengthened by the plot in the Figure~\ref{fig:sq_diff}, where we show the normalized sum of squared difference of synaptic weights between perturbed and unperturbed systems and find that to constantly increase. 

\subsection{Brief exposition of our result}
We now provide a brief exposition of why the dynamics governed by the STDP update rule might be sensitive to small perturbations to synaptic weights. The formal analysis is described in detail in the later sections. 
Let us consider the two events - \emph{birth of an output spike} followed by a \emph{birth of an input spike} to happen after a tiny increment of weight in one of the input synapses. Due to larger weight in the perturbed synapse, the perturbed neuron will spike little earlier compared to the unperturbed one. Hence the time difference between the current output spike and the past input spikes is smaller in the perturbed system giving rise to larger increment in the synaptic weights via the additive STDP rule. This enhances the differences in the synaptic weights. Now for an impending input spike, the most recent output spike in the perturbed system is moved little earlier to the past (as it spiked earlier). Hence the negative STDP update would be less in the perturbed system in comparison to the unperturbed one, thereby further enlarging the weight perturbation. This gives us an inkling that the additive STDP rule might be sensitive to small weight perturbations, viz. sensitive to small perturbations of input or output spike timings. 

\section{Phase space dynamics}
\label{sec:phasespace}
Before we analyze the aforementioned events in greater detail, let us consider phase space of our dynamical system.
In order to track the weight value at the synapses, we need to consider all the past input and output spikes within the neuron's synaptic efficacy window. As an input spike picks up the current weight associated with the input synapse at which it took birth, it is convenient to associate past synaptic weights with its imminent input spike. The phase-space of our dynamical system will thus consist of all past weights associated with the input spikes, current synaptic weights and the past output spikes within the synaptic efficacy window.

As the number of input and output spikes in the synaptic efficacy window generally varies with time, so will be the dimensionality of our phase space. Since the maximum number of spikes in a synapse that any neuron can have within its efficacy window $T$ is bounded by $L = \lceil\frac{T}{r}\rceil$, where  $r$ is the refractory period for that neuron, an easier fix to the changing dimensionality problem is to have our phase-space be of constant dimension $D=m+mL+L$, where $m$ is the number of input synapses. The quantity $mL$ and $L$ are for the input and output spikes respectively. If a synapse has less than $L$ spikes, the remaining entries are set to $0$. 

We study the effects of the six events, namely ($1$) \emph{birth of an output spike}, ($2$) \emph{birth of an input spike}, ($3$) \emph{death of an output spike}, ($4$) \emph{death of an input spike}, ($5$) \emph{ceiling} and ($6$) \emph{floor}. The last two events are the manifestation of the additive STDP rule. We will discuss effect of each of these events in the following subsections.
\subsection{Birth of an Output Spike}
\label{sec:birthoutspike}
Let $\Upsilon$ be the threshold for the membrane potential and let $k_{j}$ correspond to the event when the $j^{th}$ spike occurred at either the input synapse or the output synapse. As before, let $x_{i}^{k_j}$ and $x_{0}^{k_j}$ denote the time of occurrence of the $j^{th}$ spike in the $i^{th}$ input synapse and the output synapse respectively and let $w_{i}^{k_j}$ denote the weight associated with the input spike $x_{i}^{k_j}$ with values between $w_{max}$ and $w_{min}$. Let $n_i$ and $n_0$ denote the number of spikes in the $i^{th}$ input synapse and output synapse respectively within the spike efficacy window. 
Consider the event when the \emph{unperturbed} system produced an output spike where we have
\begin{eqnarray}
\label{eq:threshold}
\Upsilon &=& \sum_{i=1}^{m} \sum_{j=1}^{n_i} w_{i}^{k_{j}} P_i(x_{i}^{k_j}) + \sum_{j=1}^{n_0} P_0(x_{0}^{k_j}).
\end{eqnarray}
Let  $\Delta w_{i}^k$  denote the initial weight perturbation for the $i^{th}$ synapse just before this $k^{th}$ event. Let $\Delta x_{0}^{k}$ denote the perturbation in the output spike time due to this weight perturbation. Let the perturbations associated with the past output spikes $x_{0}^{k_j}$ be denoted by $\Delta x_{0}^{k_j}$.
We then have,
\begin{eqnarray}
\Upsilon &=& \sum_{i=1}^{m} \sum_{j=1}^{n_i} (w_i^{k_{j}} + \Delta w_i^{k_{j}}) P_i(x_{i}^{k_j}-\Delta x_{0}^{k}) + \sum_{j=1}^{n_0} P_0(x_{0}^{k_j} +\Delta x_{0}^{k_j} - \Delta x_{0}^{k}).
\end{eqnarray}
Using $1^{st}$ order Taylor expansion, equating with ($\ref{eq:threshold}$) and ignoring any second order term we have,
\begin{equation}
\Upsilon = \Upsilon +  \sum_{i=1}^{m} \sum_{j=1}^{n_i} \Delta w_i^k P_i(x_{i}^{k_j}) + \sum_{j=1}^{n_{0}} \frac{\partial P_{0}}{\partial x_{0}^{k_j}}(\Delta x_{0}^{k_j} ) - \Delta x_{0}^{k} \left( \sum_{i=1}^{m} \sum_{j=1}^{n_i} \frac{\partial P_{i}}{\partial x_{i}^{k_j}} w_i^{k_{j}} + \sum_{j=1}^{n_{0}} \frac{\partial P_{0}}{\partial x_{0}^{k_j}}\right ). \nonumber
\end{equation}
Hence the resulting output perturbation because of changes in the synaptic weight is given by,
\begin{equation}
\label{eq:deltay0}
\Delta x_{0}^{k} = \frac{\sum_{i=1}^{m} \sum_{j=1}^{n_i} \Delta w_i^{k_{j}} P_i(x_{i}^{k_j}) + \sum_{j=1}^{n_{0}} \frac{\partial P_{0}}{\partial x_{0}^{k_j}}(\Delta x_{0}^{k_j} )}{\lambda}
\end{equation}
where, 
\begin{equation}
\label{eq:lambda}
\lambda = \sum_{i=1}^{m}\sum_{j=1}^{n_{i}} \frac{\partial P_{i}}{\partial x_{i}^{k_j}} w_i^{k_{j}}  + \sum_{j=1}^{n_{0}} \frac{\partial P_{0}}{\partial x_{0}^{k_j}}.
\end{equation}
\newline
From the additive STDP update, an output spike increases the $i^{th}$ synapse's weight by 
\newline
$\xi_i^k = w_{max}\sum_{j=1}^{n_i} A_{+} e^{\frac{- x_{i}^{k_j}}{\tau_{+}}}$ where $A_{+}$, $\tau_{+}$ are constants for positive update of additive STDP. 
The new updated weight for $i^{th}$ synapse after this event equals $w_{i}^{k+1} = w_{i}^k+\xi_{i}^k$.

Now consider the perturbed system. Let $\tilde{w_{i}}^k$ denote the perturbed weight at the $i^{th}$ synapse before the occurrence of this output spike, i.e, $\tilde{w_{i}}^k = w_{i}^k + \Delta w_{i}^k$. Let $\tilde{w_{i}}^{k+1}$ denote the new weight in the perturbed system \emph{after} the $k^{th}$ event. Then from STDP rule we have,
\begin{equation}
\tilde{w_{i}}^{k+1} = \tilde{w_{i}}^k + w_{max}\sum_{j=1}^{n_i} A_{+} e^{\frac{-(x_{i}^{k_j}-\Delta x_0^k)}{\tau_+}}.\nonumber
\end{equation}
Using the $1^{st}$ order Taylor expansion and plugging the value of $\tilde{w_{i}}^k$ we get,
\begin{equation}
\label{eq:pertb}
\Delta w_i^{k+1} = \tilde{w_{i}}^{k+1}- w_{i}^{k+1}= \Delta w_i^k + w_{max}\sum_{j=1}^{n_i} A_{+} \, e^{\frac{- x_{i}^{k_j}}{\tau_{+}}} \frac{\Delta x_{0}^{k}}{\tau_{+}}. \nonumber
\end{equation}
Substituting the value of $\Delta x_{0}^{k}$ from Equation~\ref{eq:deltay0} we can write
\begin{eqnarray}
\label{eq:birthoutspike}
\Delta w_i^{k+1} &=& \Delta w_{i}^k + \left( \sum_{j=1}^{n_i} A_{+} \, e^{\frac{- x_{i}^{k_j}}{\tau_{+}}} \right) \frac{w_{max}}{\tau_{+}} \left( \frac{\sum_{i=1}^{m} \sum_{j=1}^{n_i} \Delta w_{i}^{k_{j}} P_i(x_{i}^{k_j}) + \sum_{j=1}^{n_{0}} \frac{\partial P_{0}}{\partial x_{0}^{k_j}}\Delta x_{0}^{k_j} }{\lambda} \right). \nonumber \\
\end{eqnarray}

\begin{figure}[ht!]
	\centering
    \includegraphics[scale=0.6]{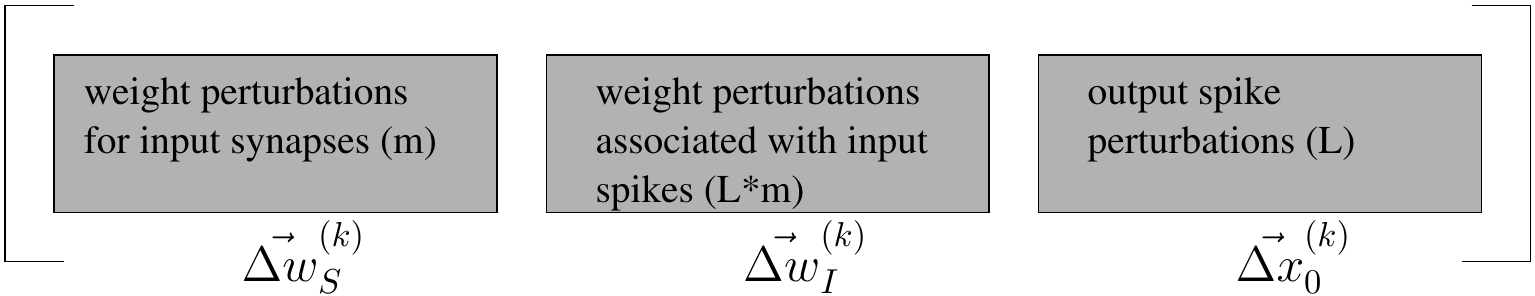}
    \caption{Components of the perturbation vector}
    \label{fig:vec}
\end{figure}

In order to analyze the sensitivity of the STDP rule, it suffices to study the dynamics of the perturbation vector during the occurrence of these events. As briefly mentioned before, the perturbation vector has three components - (i) weight perturbations associated with $m$ synapses (${\vec{\Delta w}}_S^{(k)}$)
(ii) weight perturbations associated with all the input spikes (${\vec{\Delta w}}_I^{(k)}$) (iii) output spikes perturbations(${\vec{\Delta x}}_0^{(k)}$). Figure~\ref{fig:vec} shows all the components with appropriate dimensions. 

Let
\begin{equation}
\label{eq:uvz}
u_i^k = \frac{w_{max}}{\tau_{+}} \sum_{j=1}^{n_i} A_{+} \, e^{\frac{ -x_{i}^{k_j}}{\tau_{+}}} \quad \mbox{and} \quad
v_{i}^{k_j} = \frac{ P_i(x_{i}^{k_j})} {\lambda} \quad \mbox {and} \quad
z_{0}^{k_j} = \frac{1}{\lambda} \frac{\partial P_{0}}{\partial x_{0}^{k_j}}.
\end{equation}
Define the vectors
\begin{eqnarray}
\label{eq:u_k}
\vec u^{(k)} &=& [ u_{1}^{k}, u_{2}^{k} \cdots, u_{m}^k], \hspace{10pt} \vec{z_0}^{(k)} = [{z_0}^{k_1}, {z_0}^{k_2}, \cdots, {z_0}^{k_{n_0}}, 0, 0, \cdots 0], \nonumber \\
\label{eq:v_k}
\vec v^{(k)} &=& [ v_{1}^{k_{1}},\cdots, v_{1}^{k_{n_1}},0,\cdots,v_{2}^{k_1},\cdots,v_{2}^{k_{n_2}},0,\cdots,v_{m}^{k_{1}}, \cdots, v_{m}^{k_{n_m}},0, \cdots] \nonumber
\end{eqnarray}
and the perturbation vectors
\begin{eqnarray}
\label{eq:del_w_s}
\vec {\Delta w}_{S}^{(k)} &=& [ \Delta w_{1}^{k}, \cdots, \Delta w_{m}^{k} ], \hspace{10pt}  \vec{\Delta x_0}^{(k)} = [{\Delta x_{0}}^{k_1},\cdots, {\Delta x_{0}}^{k_{n_0}}, 0, 0, \cdots, 0] \mbox{  and  }\nonumber \\
\label{eq:delta_w_k}
\vec{\Delta w}_{I}^{(k)} &=& [ \Delta w_{1}^{k_1},\cdots, \Delta w_{1}^{k_{n_1}}, 0, \cdots, \Delta w_{2}^{k_1},\cdots, \Delta w_{2}^{k_{n_2}},0, \cdots, \Delta w_{m}^{k_1}, \cdots, \Delta w_{m}^{k_{n_m}},0,\cdots]. \nonumber
\end{eqnarray}
Note that in the above definition, some of the vectors have zeros appended to it in order to maintain the constant dimensionality of the phase-space as described under Section~\ref{sec:phasespace}. Weight update takes place only at the synapses while the weights associated with input synapses remain intact. The new output perturbation $\Delta x_{0}^{k}$ gets added to the output perturbation vector $\vec{\Delta x}_0^{(k)}$. The matrix governing the dynamics of the perturbation vector is shown in Figure~\ref{fig:matrix_1} where $I$ represents the identity matrix of appropriate dimension and $J_1$ is an identity matrix appended with a last column of all zeros.

Firstly note that $u_i^k$ is positive as it is sum of exponential each multiplied by a positive constant $A_{+}$.
For excitatory synapses, the numerator term in the definition of $v_{i}^{k_j}$ namely, $P_i(x_{i,j})$, is positive as it is just the post synaptic potential (PSP) corresponding to the input spike $x_{i,j}$. For similar reasons, the numerator will be negative if the $i^{th}$ synapse is an inhibitory one. Since STDP rule doesn't apply for inhibitory synapses (as their synaptic strength doesn't change) we may set $v_{i}^{k_j}=0, \forall j$ as its corresponding $\Delta w_{i}^{k_j}=0$\,-- the latter is true as there is no perturbation at an inhibitory synapse before the $k^{th}$ event. In other words, as $v_{i}^{k_j}\Delta w_i^{k_j}=0$ \emph{independent} of $v_{i}^{k_j}$, without loss of generality we can set $v_{i}^{k_j}=0$. Additionally $u_i^k=0$ for an inhibitory synapse. So we can restrict ourselves only to excitatory synapses in our subsequent analyses.

To see $v_{i}^{k_j} > 0$ for an excitatory synapse, observe that its denominator $\lambda$ is the derivative of the membrane potential at the time of the output spike in the unperturbed system and hence is positive. Also the components of the $\vec{z_0}^{(k)}$ namely $z_0^{k_j}$ are strictly positive as its numerator is just the derivative of the AHP which is a strictly increasing function. Hence we notice that all the entries in the perturbation matrix are \emph{non-negative}, a fact that is used in Section~\ref{sec:sensitivityanalysis} to show the sensitivity of the additive STDP rule.
 
\subsection{Birth of an Input Spike}
\label{sec:birthinspike}
\begin{figure}[ht]
\begin{minipage}[b]{0.5\linewidth}
\centering
\includegraphics[scale=0.37]{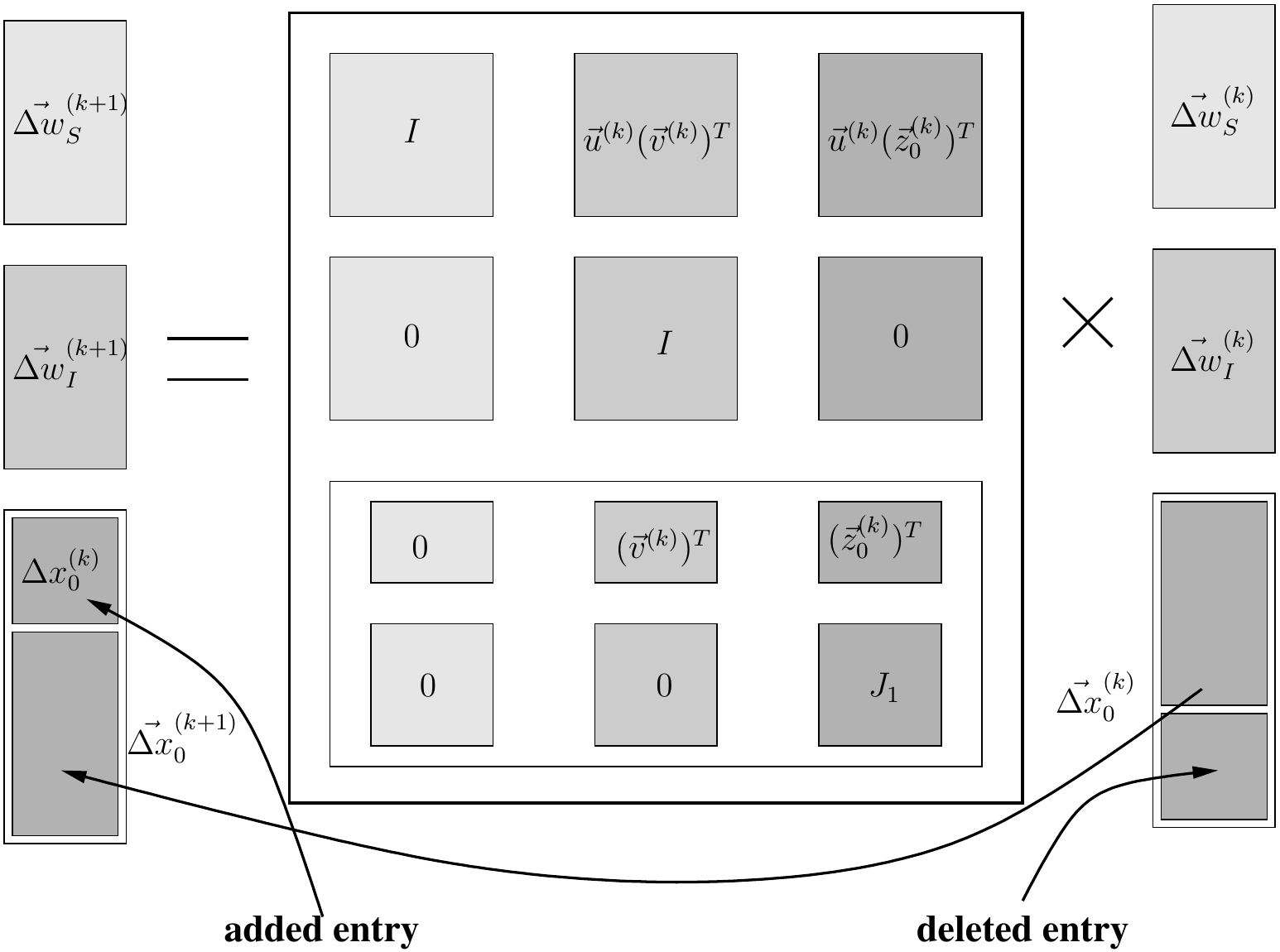}
\caption{Matrix for birth of an output spike}
\label{fig:matrix_1}
\end{minipage}
\hspace{0.5cm}
\begin{minipage}[b]{0.5\linewidth}
\centering
\includegraphics[scale=0.37]{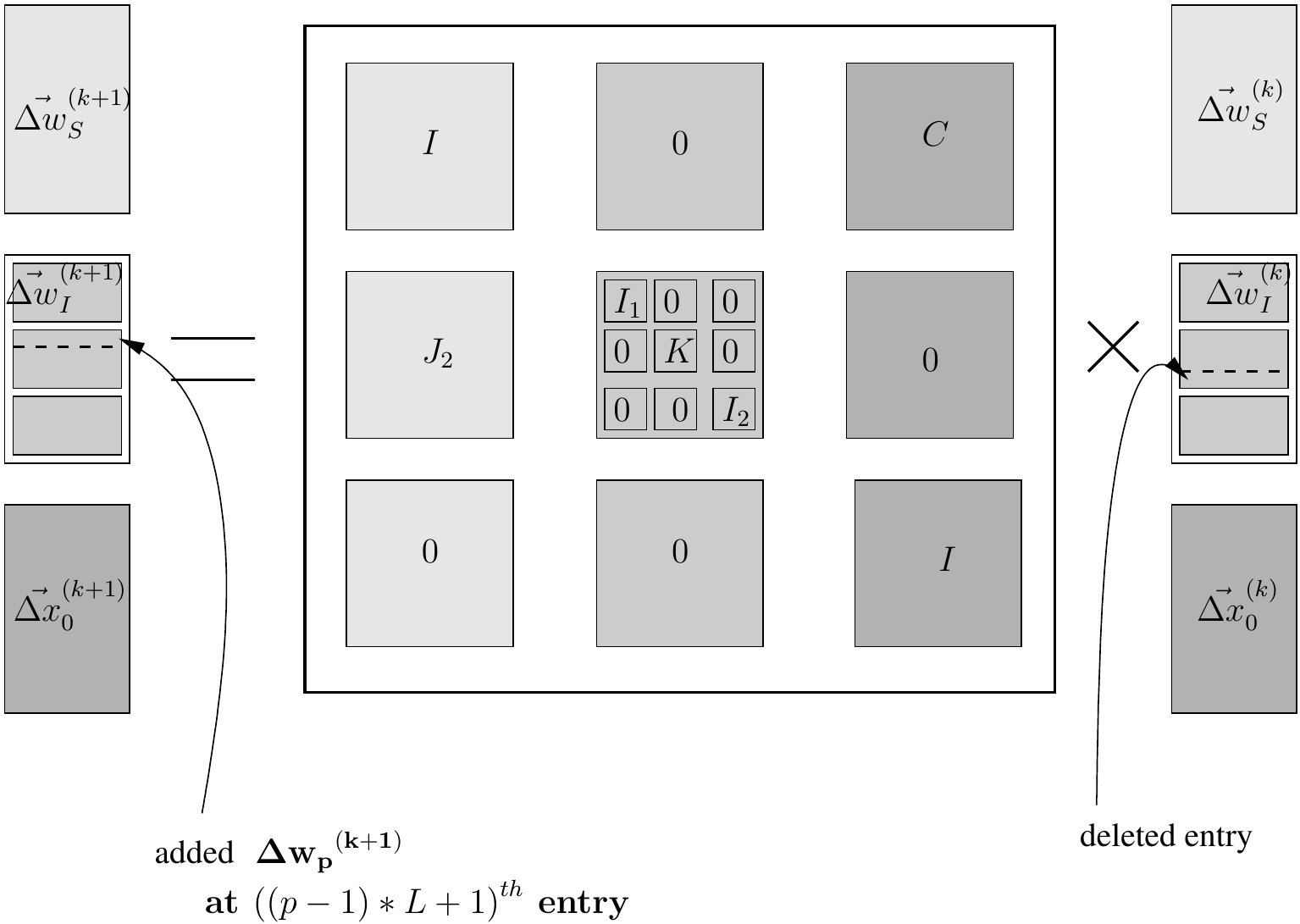}
\caption{Matrix for birth of an input spike}
\label{fig:matrix_2}
\end{minipage}
\end{figure}

For our analysis we consider the event \emph{birth of a input spike} to have occurred \emph{after} a set of past output spikes in the neuron's STDP efficacy window. Hence an input spike will decrease the synaptic weight at the $p^{th}$ synapse where it took birth. The decrement based on additive STDP rule is given by $\xi_p^k = w_{max}\sum_{j=1}^{n_0} A_{-} e^{\frac{-x_{0}^{k_j}}{\tau_{-}}}$ where $x_{0,j}$ denotes the past output spike timings, $A_{-}$ and $\tau_{-}$ are positive constants for negative STDP update. Closely following the steps delineated in Section~\ref{sec:birthoutspike} and using the $1^{st}$ order Taylor expansion, the synaptic weight difference at the $p^{th}$ synapse between the perturbed and the unperturbed systems after the arrival of the current input spike can be computed as
\begin{equation}
\label{eq:birthinpspike}
{\Delta w_p}^{k+1} = {\Delta w_p}^k + w_{max}\sum_{j=1}^{n_0} A_{-} \, e^{\frac{-x_{0}^{k_j}}{\tau_{-}}} \frac{\Delta x_{0}^{k_{j}}}{\tau_{-}}.
\end{equation}
Moreover, as the newly born input spike will pick up the previous synaptic weight and its corresponding weight perturbation from the $p^{th}$ synapse, a new entry needs to be added to $\vec{\Delta w_I}$ at the $(p-1)L+1$ location with a value of $\Delta w_p^{k}$. The perturbation matrix is given in Figure~\ref{fig:matrix_2}.
$J_2$ is a zero matrix except having $1$ at the ($(p-1)L+1,p$) entry for addition of $\Delta w_{p}^{k}$ element
to $\vec{\Delta w_I}$ at the $(p-1)L+1$ location. Both $I_1$ and $I_2$ are identity matrices of dimension $(p-1)L$ and $(m-p)L$ respectively corresponding to unaltered weight perturbations associated with input spikes in other synapses. $K$ is another identity matrix appended with a first row and last column of all zeros. As all the synaptic weights except for the $p^{th}$ synapse remain unscathed, all rows in $C$ barring the $p^{th}$ row are zero and the entries in the $p^{th}$ row are given by $C_{p_{j}}= \frac{w_{max}}{\tau_{-}}A_{-} \, e^{\frac{-x_{0}^{k_j}}{\tau_{-}}}$. Importantly notice that all entries in the perturbation matrix are again non-negative. We will make use of this fact in our subsequent analysis.

\subsection{Death of an output and input spike}
\begin{figure}[ht]
\begin{minipage}[b]{0.5\linewidth}
\centering
\includegraphics[scale=0.37]{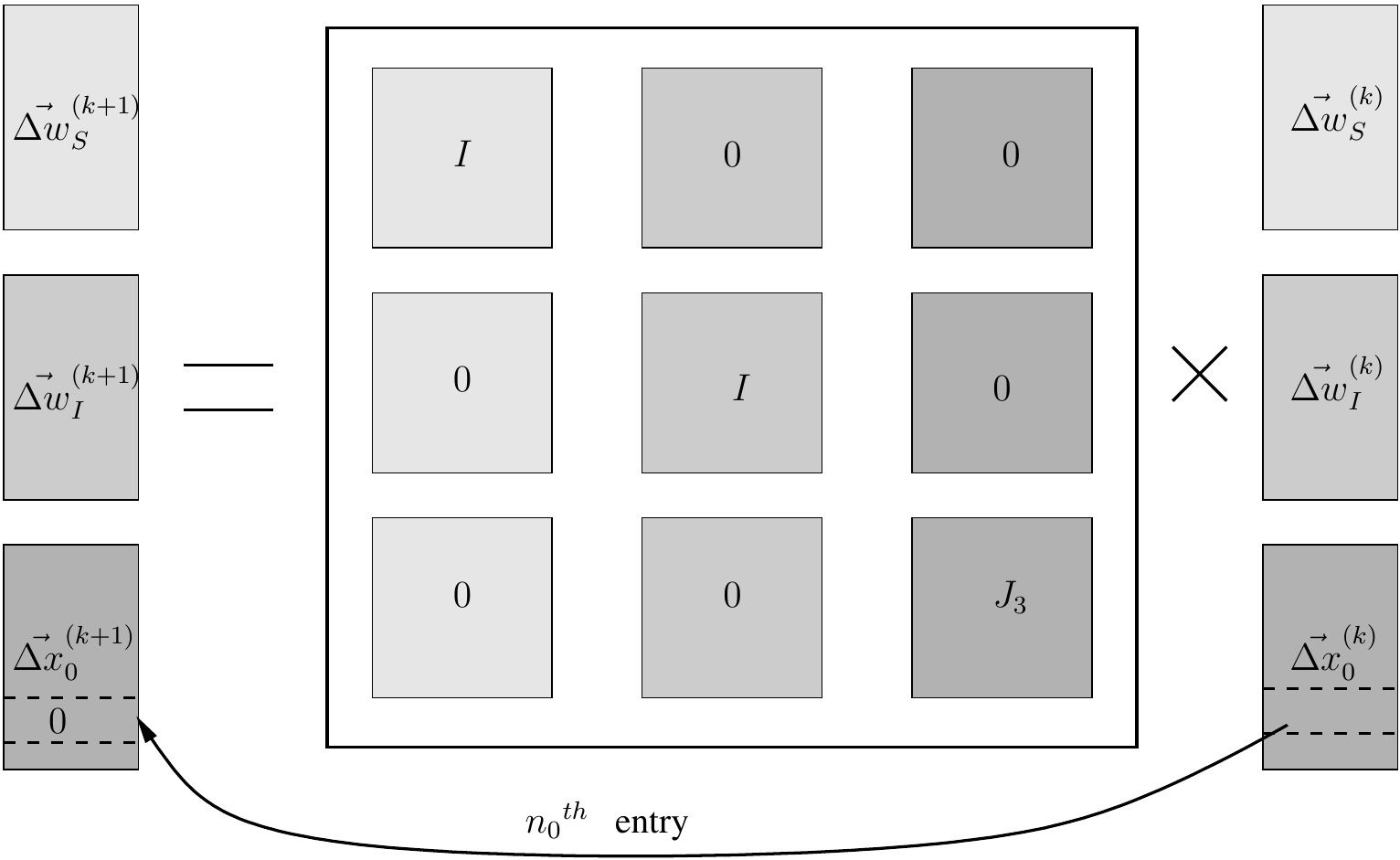}
\caption{Matrix for death of an output spike}
\label{fig:matrix_3}
\end{minipage}
\hspace{0.5cm}
\begin{minipage}[b]{0.5\linewidth}
\centering
\includegraphics[scale=0.37]{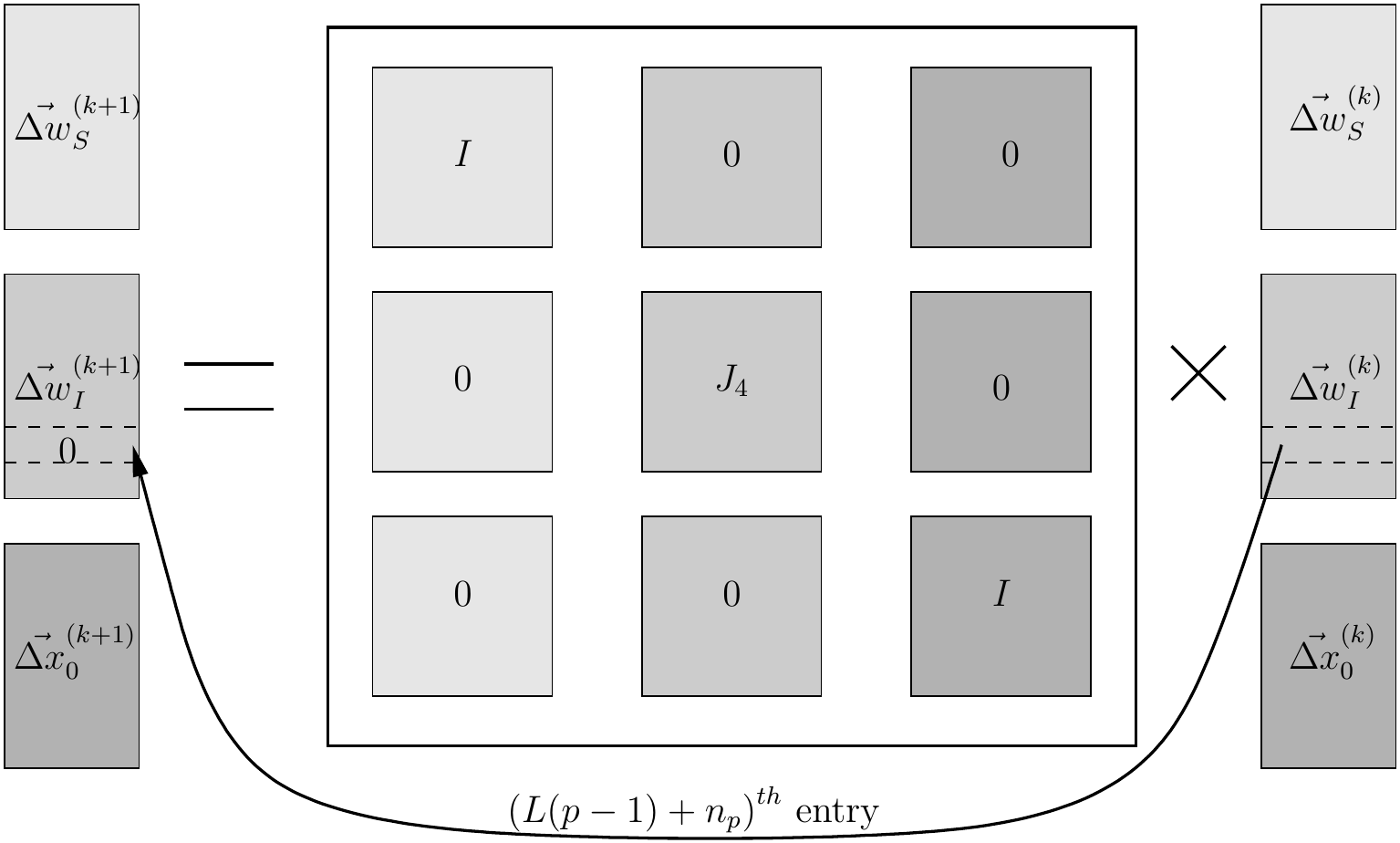}
\caption{Matrix for death of an input spike}
\label{fig:matrix_4}
\end{minipage}
\end{figure}

These events do not change the synaptic weights. Only the entries in the $n_0^{th}$ and $(L(p-1)+n_p)^{th}$ locations in $\vec{\Delta x}_0^{(k)}$ and $\vec{\Delta w}_I^{(k)}$ are set to $0$ respectively. $p$ denotes the synapse where the event \emph{death of an input spike} took place. The perturbation matrices for these events are shown in Figures~\ref{fig:matrix_3} and \ref{fig:matrix_4}. Here, $I$ denotes the identity matrix of appropriate dimension. $J_3$ and $J_4$ are very similar to the identity matrix except they have one row of all zeros for setting the corresponding entries to $0$ in $\vec{\Delta x}_0^{(k)}$ and $\vec{\Delta w}_I^{(k)}$.

\subsection{Ceiling and Floor Events}
It is well known in literature \cite{song2000competitive} that when the dynamics of the synaptic weights are governed by the additive STDP rule, an artificial cap needs to be imposed on the maximum (minimum) values that the weights can take. During the occurrence of these ceiling (floor) events corresponding to the weights attaining the maximum (minimum) values, the corresponding STDP rule is not applied and the weights values stays put. As a result, the entries of $\Delta w_S$ corresponding to the synapses which faces these events will be set to zero in the worst case nullifying the perturbation. 

\section{Sensitivity Analysis}
\label{sec:sensitivityanalysis}
We begin our sensitivity analysis by demonstrating that except for the ceiling and the floor events, the additive STDP rule is sensitive to small synaptic weight perturbations. Since the occurrence of the ceiling and the floor events are stochastic in nature and completely depends on the input spike distribution, the neutralizing effect of these events on the perturbation vector needs to be modeled via a stochastic process as described in the subsequent section. The overall sensitivity of the STDP rule will depend
on the interplay between the enhancing effects of the birth and death events and the squashing effect of the ceiling and floor events on the perturbation vector.

Let $A^{(k)}$ denote the perturbation matrix for the $k^{th}$ event and let $\vec{\chi}^{(k+1)}$ represent the perturbation vector after the $k^{th}$ event, i.e., $\vec{\chi}^{(k+1)} = A^{(k)} \vec{\chi}^{(k)}$. It is worth recalling that all entries of $A^{(k)}$ are \emph{non-negative} irrespective of the type of event. Additionally the perturbation vector $\vec{\chi}^{(k)}$ is made of three components namely $\chi^{(k)} = [\vec{\Delta w_S}^{(k)} \vec{\Delta w_I}^{(k)} \vec{\Delta x}^{(k)}]$
as shown in Figure~\ref{fig:vec}. 

Let $\mathbb{A}^{(n)} = A^{(n)}*\cdots * A^{(1)}$ denote the combined effect of these perturbation matrices after the occurrence of $n$ events. In order to prove the sensitivity of the additive STDP to small perturbation, it suffices to show that as $n \rightarrow \infty$, the Frobenius norm of the matrix $\mathbb{A}^{(n)}$ goes to infinity. In order to show this, without loss of generality we take a specific initial perturbation vector $\vec{\chi}^{(1)}$ given by $ \vec{\chi}^{(1)} = \left[ \left \{ 0,\cdots,0,\Delta w_{p}^{1},0,\cdots,0 \right \}, \vec{0}, \vec{0} \right]$
where only the $p^{th}$ synapse (for any $p$) has a non-zero positive perturbation $\Delta w_{p}^{1}$. Both the input weight perturbation $\left(\vec{\Delta w_I}^{(1)}\right)$ and the output perturbation $\left(\vec{\Delta x}^{(1)}\right)$ vectors are set to $\vec{0}$. Then $\|\vec{\chi}^{(1)}\|_2 = \|\vec{\Delta w_S}^{(1)}\|_2 =  \Delta w_{p}^{1}$. We now show that this choice of $\vec{\chi}^{(1)}$ works, i.e.,
$\|\vec{\chi}^{(n+1)}\|_2 \geq \|\vec{\Delta w_S}^{(n+1)}\|_2 > \|\vec{\Delta w_S}^{(1)}\|_2 = \|\vec{\chi}^{(1)}\|_2$,
for sufficiently large $n$. Specifically we prove that the norm of $\vec{\Delta w}_S^{(k)}$ \emph{always} increases with the occurrence of specific type of events and is \emph{never} decreased by any of these events (except for the \emph{ceiling} and \emph{floor} events for which we gave the stochastic analysis in the next section).

Since all the entries of $A^{(k)}$ are \emph{non-negative}, it is easy to see that all entries of $\vec{\chi}^{(k+1)}$ are also non-negative even after the occurrence of any of these events. Focusing on the dynamics of the first component of $\vec{\chi}^{(k+1)}$ namely $\vec{\Delta w_S}^{(k+1)}$, firstly, observe that the occurrence of death of either an output spike or an input spike doesn't affect $\vec{\Delta w_S}^{(k+1)}$ as the synaptic weights remain unruffled. Secondly the birth of the output or input spike always increases the norm of $\vec{\Delta w_S}^{(k+1)}$ as either all its components (in case of birth of an output spike) or one of its components (in case of birth of an input spike) gets a positive increment. This is easily verifiable from Equations~\ref{eq:birthoutspike} and \ref{eq:birthinpspike} because of positivity of $\vec{\chi}^{(k)}$. In fact we can claim that this positive increments are at least $\epsilon$ which are bounded away from $0$. In short, the $\|\vec{\Delta w_S}^{(k)}\|_2$ increases with the occurrence of birth of either the output or input spikes and doesn't change with the death of these spikes. Hence $\|\vec{\chi}^{(n+1)}\|_2 > \|\vec{\chi}^{(1)}\|_2$ for sufficiently large $n$ with increments bounded away from 0, which in turn suggests that the Frobenius norm of $\mathbb{A}^{(n)}$ will increase and converge towards \emph{infinity}. 

\subsection{Characterizing the Underlying Stochastic Process}
The aforementioned analysis requires an investigator who has access to the system of this spiking neuron from which he draws data to compute the perturbation matrix $A^{(k)}$ for all the birth, death, ceiling and floor events. With the absence of the last two events, our neural system is sensitive to small weight perturbations irrespective of whatever statistics it might have. The occurrence of the ceiling and floor events and its stochastic nature warrants the development of a stochastic model for our analysis. We assume that both the input and output spike distribution is a homogeneous Poisson process with the mean rate of $\lambda_{in}$ and $\lambda_{out}$ respectively.  Let $p_1$, $p_2$, $p_3$, $p_4$ be the steady state probability of occurrence of birth of an output spike, birth of an input spike, death of an output spike and death of an input spike respectively. Since the occurrence of the ceiling event requires that the synaptic weights attain or exceeds its maximum value, it can take birth only after the birth of an output spike. Similarly, the floor event can occur only after the birth of an input spike. Let $\gamma_1$ and $\gamma_2$ be the probability of occurrence of ceiling and floor events given the occurrence of the birth of an output and input spike respectively and also assume that with probability $\gamma_3$, a synapse individually experiences the ceiling event given that the ceiling event has already occurred at that time point. We now describe the stochastic process for the generation of the perturbation matrices $A^{(k)}$ corresponding to each of these events. The process is begun by choosing integers $n_i (i=1,\cdots, m)$ and $n_0$ randomly over the range $[0,L]$ to denote number of input spikes in the $i^{th}$ synapse and output spikes inside the synaptic efficacy window respectively. 

In case of birth of an output spike, the perturbation matrix $A^{(k)}$ is shown in Figure~\ref{fig:matrix_1}. The entries of the matrix are functions of $u_i^k$, $v_i^{k_j}$ and $z_0^{k_j}$ given in Equation~\ref{eq:uvz} which in turn depends on the input and output spikes--through the PSP and AHP functions and its derivatives--and the weights associated with the input spike train. Hence the matrix can be populated by sampling from the input and output spike times which follow Poisson distribution and from the synaptic weight distribution which is known to be bimodal \cite{song2000competitive} and then computing the $u_i^k$, $v_i^{k_j}$ and $z_0^{k_j}$ values. Care should be taken to ensure that the value of $\lambda$ given in Equation~\ref{eq:lambda} is positive. We also assume that the choice of birth and death events are \emph{independent} of the entries of the matrix. In other words the distribution of the entries obtained as function of input and output spikes and synaptic weights is independent of the birth and death events. If the \emph{ceiling} event follows this output event (with the probability of $\gamma_1$), then all the entries of certain rows in the constructed perturbation matrices will be set to zero. 


In case of birth of an input spike, the choice of the synaptic index $p$ at which the event is expected to occur is randomly chosen between 1 to $m$ ($m$ is the number of input synapses) \emph{independent} of the previous events.
Once the choice is made, the corresponding perturbation matrix $A^{(k)}$ shown in Figure~\ref{fig:matrix_2} is populated 
by sampling from the output spike time Poisson distribution and then computing the $C_{p_j}$ values. The distribution of $C_{p_j}$ values obtained as a function of output spike train distribution is assumed to be independent of the past events. If the \emph{floor} event follows this input event (with probability $\gamma_2$), then all entries of $p^{th}$ row of $A^{(k)}$ are set to zero.

In case of a death of an output event, the perturbation matrix is an identity matrix except that the entries of the row corresponding to the last existing output spike is set to zero. Similarly, in case of a death of an input event, the synaptic index $p$ at which the death event is expected to occur is again chosen randomly between 1 to $m$. The perturbation matrix is an identity matrix except that the entries of the row corresponding to the last existing spike in the $p^{th}$ synapse is set to zero.


\begin{figure}
  \begin{center}
    \includegraphics[scale=0.35]{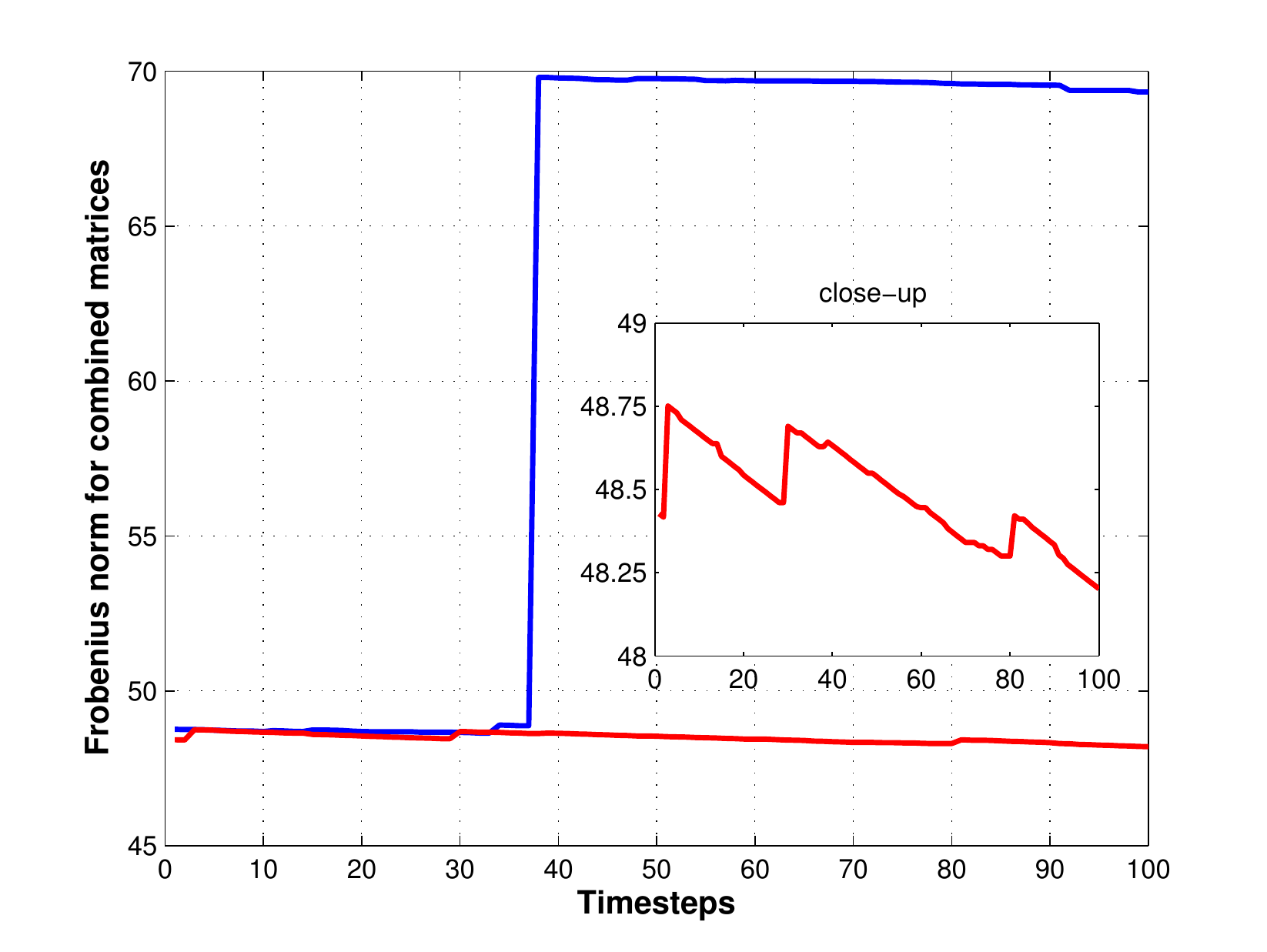}
  \end{center}
  \caption{Frobenius norm of $\mathbb{A}^{(n)}$. (i) Blue: High ceiling and floor events, (ii) Red: Low ceiling and floor events}
  \label{fig:F_norm_both}	
\end{figure}

In Figure~{\ref{fig:F_norm_both}} on left, two very different scenarios of simulation are presented. Both curves indicate the Frobenius norm of $\mathbb{A}^{(n)}$ computed in various timesteps. The blue curve shows the rise of the Frobenius norm, which clearly reaffirms our analysis as we have lesser probabilities for \emph{ceiling} and \emph{floor} events ($\gamma_1=\gamma_2=0.1$) and lesser probability for \emph{ceiling} event for each synapse ($\gamma_3 = 0.1$) as well. On the other hand, the red curve shows minimum variance with a decreasing trend as we have $\gamma_1=\gamma_2=\gamma_3=0.9$, which decreases the Frobenius norm by generating more and more \emph{ceiling} and \emph{floor} events.

\section{Discussion}
We studied the dynamics of the synaptic weights of a single neuron where the weights are modified according to the additive STDP rule. Interestingly, we noticed that the individual synaptic weight values between the perturbed and the unperturbed system tend to vary significantly as they evolve in time demonstrating the sensitivity of the additive STDP rule to tiny perturbations. If either the synaptic weight, or the input and output spike times of a neuron were to be infinitesimally perturbed at a given point in time, additive STDP would cause the neuron to reach vastly different states, even when the subsequent input spike train is identical.
 We also noticed that as the frequency of the ceiling and floor events are increased (when the weights reach their maximum or minimum values respectively)---scenarios where the STDP rule is not applied---the system is more or less stable and becomes \emph{insensitive} to perturbations which is quite counter-intuitive.  
In the future, we would like to investigate other kinds of STDP rule like multiplicative, add-multiplicative, mixed STDP rules. We would like to recover the criterion which decides the sensitivity of the system. In a seminal article \cite{PhysRevLett.64.1196}, it has been shown how one can convert a chaotic attractor to any one of a large number of possible attracting time-periodic motions by making only small time-dependent perturbations of an available system parameter. Another fruitful research prospect would be to study how to counter-balance sensitivity effects and drive the system to the desired state.

\newpage
\clearpage
\noindent 
\bibliographystyle{siam}
\bibliography{stdp_sensitivity}

\begin{thebibliography}{10}

\bibitem{abbott2004homeostasis}
{\sc L.F. Abbott and W.~Gerstner}, {\em Homeostasis and learning through
  spike-timing dependent plasticity}, in Methods and models in Neurophysics,
  C.~Hansel, D.~Chow, B.~Gutkin, and C.~Meunier, eds., 2004.

\bibitem{abbott2000synaptic}
{\sc L.F. Abbott and S.B. Nelson}, {\em Synaptic plasticity: {T}aming the
  beast}, Nature Neuroscience, 3 (2000), pp.~1178--1183.

\bibitem{banerjee2006sensitive}
{\sc A.~Banerjee}, {\em On the sensitive dependence on initial conditions of
  the dynamics of networks of spiking neurons}, Journal of Computational
  Neuroscience, 20 (2006), pp.~321--348.

\bibitem{bi1998synaptic}
{\sc G.Q. Bi and M.M. Poo}, {\em Synaptic modifications in cultured hippocampal
  neurons: {D}ependence on spike timing, synaptic strength, and postsynaptic
  cell type}, The Journal of Neuroscience, 18 (1998), pp.~10464--10472.

\bibitem{caporale2008spike}
{\sc N.~Caporale and Y.~Dan}, {\em Spike timing-dependent plasticity: {A}
  {H}ebbian learning rule}, The Annual Review of Neuroscience, 31 (2008),
  pp.~25--46.

\bibitem{dan2004spike}
{\sc Y.~Dan and M.M. Poo}, {\em Spike timing-dependent plasticity of neural
  circuits}, Neuron, 44 (2004), pp.~23--30.

\bibitem{gerstner1996neuronal}
{\sc W.~Gerstner, R.~Kempter, J.L. van Hemmen, and H.~Wagner}, {\em A neuronal
  learning rule for sub-millisecond temporal coding}, Nature, 383 (1996),
  pp.~76--78.

\bibitem{gerstner2002spiking}
{\sc W.~Gerstner and W.M. Kistler}, {\em Spiking neuron models: {S}ingle
  neurons, populations, plasticity}, Cambridge Univ Press, 2002.

\bibitem{karmarkar2002model}
{\sc U.R. Karmarkar and D.V. Buonomano}, {\em A model of spike-timing dependent
  plasticity: {O}ne or two coincidence detectors?}, Journal of Neurophysiology,
  88 (2002), pp.~507--513.

\bibitem{lengyel2005matching}
{\sc M.~Lengyel, J.~Kwag, O.~Paulsen, and P.~Dayan}, {\em Matching storage and
  recall: {H}ippocampal spike timing-dependent plasticity and phase response
  curves}, Nature Neuroscience, 8 (2005), pp.~1677--1683.

\bibitem{markram1997regulation}
{\sc H.~Markram, J.~L{\"u}bke, M.~Frotscher, and B.~Sakmann}, {\em Regulation
  of synaptic efficacy by coincidence of postsynaptic aps and epsps}, Science,
  275 (1997), pp.~213--215.

\bibitem{PhysRevLett.64.1196}
{\sc E.~Ott, C.~Grebogi, and J.A. Yorke}, {\em Controlling chaos}, Physical
  Review Letters, 64 (1990), pp.~1196--1199.

\bibitem{pfister2006optimal}
{\sc J.P. Pfister, T.~Toyoizumi, D.~Barber, and W.~Gerstner}, {\em Optimal
  spike-timing-dependent plasticity for precise action potential firing in
  supervised learning}, Neural Computation, 18 (2006), pp.~1318--1348.

\bibitem{rubin2001equilibrium}
{\sc J.~Rubin, D.D. Lee, and H.~Sompolinsky}, {\em Equilibrium properties of
  temporally asymmetric {H}ebbian plasticity}, Physical Review Letters, 86
  (2001), pp.~364--367.

\bibitem{song2000competitive}
{\sc S.~Song, K.D. Miller, and L.F. Abbott}, {\em Competitive {H}ebbian
  learning through spike-timing-dependent synaptic plasticity}, Nature
  Neuroscience, 3 (2000), pp.~919--926.

\bibitem{kepecs2002neuronal}
{\sc J.~Tegner and A.~Kepecs}, {\em Why neuronal dynamics should control
  synaptic learning rules}, in Proceedings of the 2001 Neural Information
  Processing Systems (NIPS) Conference, The MIT Press, 2001, pp.~285--292.

\bibitem{van2000stable}
{\sc M.C.W. van Rossum, G.Q. Bi, and G.G. Turrigiano}, {\em Stable hebbian
  learning from spike timing-dependent plasticity}, The Journal of
  Neuroscience, 20 (2000), p.~8812.

\end{thebibliography}

\end{document}